\documentclass[11pt,twoside]{article}

\usepackage{asp2006}
\usepackage{epsf}
\usepackage{psfig}
\usepackage{lscape}
\usepackage{graphicx}
\usepackage{subfigure}
\usepackage{amsmath}
\usepackage{amssymb}

\begin{document}
\title{Are There Meteors Originated from Near Earth Asteroid (25143) 
Itokawa?}

\author{K. Ohtsuka $^{1}$, S. Abe $^{2}$, 
M. Abe $^{3}$,  H. Yano $^{3}$, J. Watanabe $^{4}$}

\affil{
$^{1}$ {Tokyo Meteor Network, 1--27--5 Daisawa, Setagaya-ku, Tokyo 155--0032, Japan}\\
$^{2}$ {Graduate School of Science and Technology, Kobe University, Nada, Kobe 
657--8501, Japan}\\
$^{3}$ {JAXA/ISAS, 3--1--1 Yoshinodai, Sagamihara 229-8510, Japan}\\
$^{4}$ {National Astronomical Observatory, Osawa, Mitaka, Tokyo 181--8588, Japan}
}

\begin{abstract}
As a result of a survey of Itokawid meteors (i.e., 
meteors originated from Near Earth Asteroid (25143) Itokawa = 1998 $\mathrm{SF_{36}}$), 
from among the multi-station optical meteor orbit data of $\sim 15000$ orbits, 
and applying the $D$-criteria, we could find five Itokawid meteor candidates. 
We also analyzed corresponding mineral materials of the Itokawid candidates 
through their trajectory and atmospheric data. We conclude, on the basis of 
our investigation, that the fireball, MORP172, is the strongest Itokawid 
candidate.
\end{abstract}

\section{Introduction}

The problem as to whether meteor showers associated with Near Earth 
Asteroids (NEAs) exist or not has been discussed by many authors (e.g., 
Olsson-Steel 1988; Hasegawa et al. 1992; Babadzhanov 1994). The most 
prominent case is the association between NEA (3200) Phaethon and the 
Geminid stream complex. However, it should be an exceptional case since 
Phaethon is considered to be a dormant or extinct cometary nucleus; hence 
the Geminid stream complex meteoroids were presumably released from Phaethon 
in its active cometary phase (Gustafson 1989). Meanwhile, meteoroids of 
asteroidal origin certainly exist and they are usually observed and 
recognized as meteorites, meteoritic fireballs, or high bulk-density 
meteors. They are probably produced as impact ejecta by asteroidal 
collisions. Although there is no asteroid--meteor association as strong as 
the Phaethon--Geminid complex yet, most likely the asteroidal meteor (or 
meteorite) streams indeed exist (e.g., Halliday 1987; Terentjeva {\&} 
Barabanov 2002; Spurn\'{y} et al. 2003).

The HAYABUSA (MUSES-C) sample return mission target, NEA (25143) Itokawa = 
1998 $\mathrm{SF_{36}}$, is one of the Apollo-type NEAs, whose orbital parameters at 
present epoch are perihelion distance $\sim 0.95$ AU and small eccentricity 
$\sim 0.28$ along with an inclination near the ecliptic plane at $\sim 1^\circ.7$. The 
taxonomic class of Itokawa is less the cometary nature than a typical 
S(IV)-type with rather high albedo (e.g., Binzel et al. 2001; Sekiguchi et 
al. 2003; Ishiguro et al. 2003). A meteoritic analog would correspond to a 
LL chondrite (Binzel et al. 2001). Since Itokawa is a small asteroid of 
$\sim 400$ m in mean axis, the escape speed from Itokawa will be only $\sim 25$ cm 
s$^{-1}$ (Ostro et al. 2004). Consequently, impact ejecta should easily 
escape from Itokawa toward interplanetary space and some may become 
meteoroids. The Earth approaches Itokawa's orbit to within 0.05 AU from late 
March to early July, over a span of three months, in which we would be able 
to observe some meteors originated from Itokawa (hereafter, we designate 
such a meteor as ``Itokawid") every year. If we can obtain the orbital and 
physical data of the Itokawid meteors, they must give us very important 
information when compared with the forthcoming analytical results of the 
HAYABUSA onboard scientific instruments and the sample of Itokawa.

On the basis of the above assumption, we surveyed whether there are Itokawid 
meteors or not from among the optical meteor orbit database, considering 
orbital similarity with Itokawa. As a result, we could locate Itokawid 
meteor candidates. We also analyzed the corresponding mineral materials of 
the Itokawid candidates through their trajectory and atmospheric data.

\section{Survey}

We called Japanese visual-meteor observers to watch the Itokawid meteor 
activity through the observing campaign from the 2001 season. However, no 
detection has been reported so far. This is supported by the negative 
results of the Japanese fireball network (JN) monitoring, a total of $\sim 500$ 
hr, during the 2000--2004 seasons. One of the reasons is that a predicted 
geocentric velocity for the Itokawids is very slow at only $\sim 6$ km s$^{-1}$, 
which should scatter their apparent radiants in a large celestial area 
around Leonis. This makes it difficult for visual meteor observers to 
distinguish a member of the Itokawids or not. Another reason is that the 
activity of asteroidal meteors is substantially weak. Indeed, we could not 
find any past Itokawid records in the several visual meteor catalogues. 
Thus, the Itokawid activity has never been reported by single-station visual 
meteor observers.

Unlike the visual meteor observations, we might trace Itokawid activity by 
retrieving data from the meteor orbit database recorded by multiple-station 
observations. Considering the orbital similarity with Itokawa, we surveyed 
whether there are Itokawid meteors among the multi-station optical 
(photographic and TV) meteor orbit data of $\sim 15000$ orbits. Our database 
includes: i) the IAU MDC optical meteor orbit data of $\sim 6000$ orbits as of 
August, 2001 (Lindblad 2001; Lindblad et al. 2001), based upon many 
astronomical papers that have been published (Steel 1996); ii) 6500 more 
optical meteor orbits from released catalogues on the internet websites 
(e.g. DMS additional photo and TV; MSS-WG TV; Ondrejov TV; etc.) and 
unpublished TMN photographic and JN fireball reductions; and iii) 
Super-Schmidt graphical reductions (McCrosky {\&} Posen 1961) of $\sim 2500$ 
orbits, although they were not included in the IAU MDC photographic meteor 
data (Lindblad 2001) by reason of their lower precision than other 
photographic reductions; they are comparable in quality to the TV reductions 
and hence they deserve to be retrieved here.

In investigating the orbital similarity between meteors or a comet/asteroid 
and meteors, we often use the $D$-criteria, e.g. $D_{\rm SH}$ (Southworth {\&} Hawkins 
1963); $D'$ (Drummond 1979); $D_{\rm N}$ and $D_{\rm R}$ (Valsecchi et al. 1999). 
We applied these $D$-criteria as a retrieving engine, mainly taking account of 
discrimination $D_{\rm SH} \mathbin{\lower.3ex\hbox{$\buildrel<\over 
{\smash{\scriptstyle=}\vphantom{_x}}$}} 0.15$. The traditional $D_{\rm SH}$ is 
defined by a distance of two points of two objects ($A$ and $B$) in the 
five-dimensional coordinate space systemized by the orbital elements, $e$: 
eccentricity, $q$: perihelion distance, $\omega $: argument of perihelion, 
$\mathit{\Omega}$: longitude of the ascending node, and $i$: inclination, as follows 
(Southworth {\&} Hawkins 1963):
\begin{equation}
\begin{split}
[D_{\rm SH} (A, B)]^2 = (e_A-e_B)^2+(q_A-q_B)^2+[2 \sin (I_{AB }/ 2)]^2 \\
+\{[(e_A+e_B) / 2][2 \sin (\mathit{\Pi}_{AB }/ 2)]\}^2 \nonumber
\end{split}
\end{equation}
where $I_{AB}$ is the angle between both orbital planes and $\mathit{\Pi}_{AB}$ is the 
difference between the longitudes of perihelion measured from the 
intersection of the orbits. Therefore, you can easily comprehend that the 
smaller $D_{\rm SH}$ is, the stronger the association between Itokawa and meteor will 
be.

\section{Results}

We found five Itokawid meteor candidates, as listed in Table 1, where the 
column heads from left to right are as follows: object name, observed date 
of the meteors in UT, $e$, $q$ in AU, $a$: semimajor axis in AU, the angular 
elements, $\omega $, $\mathit{\Omega}$, and $i$ in degree in J2000, $\alpha $ and 
$\delta $: right ascension and declination of the geocentric radiant in 
degree in J2000, and $V_G$: geocentric velocity in km s$^{-1}$. The orbital 
parameters of Itokawa at epoch 2004 July 14.0 TT (= JDT 2453200.5) taken from the 
NASA-JPL NEO Program website are also presented in Table 1, when mean 
anomaly was $35^\circ.690$. The equinox of angular data, except for those of the 
meteor 00504004, was transformed from B1950 to J2000.

The results of the $D$-criteria are also given in Table 2, where the column 
heads are the following: object name, $D_{\rm SH}$ between Itokawa and each meteor, 
$\lambda $ and $\beta $: longitude and latitude of the perihelion in degree 
in J2000 (main terms of $D'$), and $U$ and $\cos \theta$: unperturbed 
geocentric encounter velocity to the Earth and cosine of the angle between $U$ 
and the direction of Earth's heliocentric motion (main terms of $D_{\rm N}$ and 
$D_{\rm R}$). $U$ equals $\sqrt{3 - T_J}$, where $T_J$ is the Tisserand's invariant 
of each object with respect to Jupiter.

The meteors, H7220 and H10330, were picked out from the Super-Schmidt 
graphical reductions (McCrosky {\&} Posen 1961). $D_{\rm SH}$ between Itokawa--H7220 
is 0.05, the smallest among all the Itokawa--meteor pairs and comparable to 
the confirmed comet--meteor associations (e.g., 55P--Leonids, 103P--Perseids, 
Phaethon--Geminids; and others); thus both objects are probably associated 
with one another. Although outside the predicted months mentioned above, the 
meteor, HA10000, came from the Super-Schmidt automatic reductions (McCrosky 
{\&} Shao 1967). 

The fireball, MORP172, was multiply photographed by the 
Canadian fireball network, Meteorite Observation and Recovery Project 
(MORP), on April 11, 1975 (Halliday et al. 1996); $D_{\rm SH}$ of 0.09 between 
Itokawa--MORP172 is somewhat larger than that of Itokawa--H7220. However it is 
still within a probable association range. It is also notable that the 
secular near-invariants of MORP172 in the $U$-$\cos \theta $ coordinates match well 
with those of Itokawa. This suggests that both objects are dynamically in 
strong relationship to each other (Valsecchi et al. 1999). 
\twocolumn
%\rotatebox{90}{
\begin{table}[htdp]
%\begin{center}
\rotatebox{90}{
\begin{minipage}{\textheight}
%\centering
\caption{Itokawid meteor candidates}
\vspace{3mm}
%\tiny
\begin{tabular}{lccllllcclllll}
Object &  & date & &$\quad e$&$\quad q$&$\quad a$ &  $\quad \omega$ & $\quad \mathit{\Omega}$ & $\quad i$  & $\quad \alpha$ & $\quad \delta $ & $V_G $   \\
& Yr & M &  D (UT)& & (AU)&  (AU)\\[1pt]
\hline\hline
H7220 &1953 & Apr & 11.25 & 0.32 & 0.96 & 1.42 & 214.1 & 21.6 & 2 & 170.7 & $+13.7$ & 6.5 \\
H10330 & 1954 & Mar & 12.43 & 0.15 &  0.96 & 1.13 & 224.0 & 352.0 & 2 & 155.7 & $+25.7$ & 3.8 \\
HA10000 & 1957 & Feb & 25.35740 & 0.208 & 0.904 & 1.142 & 240.3 & 337.12 & 3.6 & 153.8 & $+31.5$ & 6.1 \\
MORP172 & 1975 & Apr & 11.244 & 0.236 & 0.970 & 1.27 & 213.5 & 21.05 & 5.1 & 182.2 & $+33.0$ & 5.6 \\
00504004 & 2000 & May & 04.68583 & 0.163 & 1.007 & 1.203 & 8.0 & 224.837 & 3.1 & 147.2 & $-22.7$ & 3.1 \\
Itokawa & &&&0.280 & 0.953 & 1.324 & 162.683 & 69.153 & 1.623 &&& 6. \\
\end{tabular}
\end{minipage}
\if0
\begin{minipage}{\textwidth}
\vspace{100mm}
\end{minipage}
\begin{minipage}{0.5\textheight}
%\centering
\caption{Results of $D$-criteria}
\vspace{3mm}
\begin{tabular}{lccccc}
Object & $D_{\rm SH}$ & $\lambda$ & $ \beta $ & $U$ &$\cos \theta $ \\ 
\hline\hline
H7220 & 0.05 & 235.7 & -1.1 & 0.20 & 0.65 \\
H10330 & 0.15 & 216.0 & -1.4 & 0.12 & 0.42 \\
HA10000 & 0.13 & 217.4 & -1.3 & 0.19 & 0.22 \\
MORP172 & 0.09 & 234.4 & -2.8 & 0.18 & 0.52 \\
00504004 & 0.15 & 232.8 & 0.4 & 0.10 & 0.79 \\
Itokawa && 231.8 & 0.5 & 0.19 &  0.55
\end{tabular}
\end{minipage}
\fi
}
%\end{center}
\label{table1}
\end{table}%
%}

\begin{table}[htdp]
%\begin{center}
\rotatebox{90}{
\begin{minipage}{\textheight}
%\centering
\caption{Results of $D$-criteria}
\vspace{3mm}
\begin{tabular}{lccccc}
Object & $D_{\rm SH}$ & $\lambda$ & $\beta$ & $U$ &$\cos \theta $ \\[1pt]
\hline\hline
H7220 & 0.05 & 235.7 & $-1.1$ & 0.20 & 0.65 \\
H10330 & 0.15 & 216.0 & $-1.4$ & 0.12 & 0.42 \\
HA10000 & 0.13 & 217.4 & $-1.3$ & 0.19 & 0.22 \\
MORP172 & 0.09 & 234.4 & $-2.8$ & 0.18 & 0.52 \\
00504004 & 0.15 & 232.8 & $+0.4$ & 0.10 & 0.79 \\
Itokawa && 231.8 & $+0.5$ & 0.19 &  0.55
\end{tabular}
\end{minipage}
}
%\end{center}
\label{table2}
\end{table}%

\onecolumn
\noindent

The TV meteor, 00504004, was recorded by the Ondrejov observatory team (Koten et al. 2003) 
on May 4, 2000; the $D_{\rm SH}$-criterion with Itokawa is just outside the 
0.15-limit, but its $\lambda $ and $\beta $ are very close to those of 
Itokawa and hence in a possible association range.

True anomaly of these meteors, at observed epoch, equals $180^\circ - \omega $ for 
the Super-Schmidt meteors and MORP172 and $-\omega $ for the meteor, 
00504004.

\section{Physical approach}

Furthermore, we approached the above subject in a physical analysis, in 
order to ascertain the association between Itokawa and the Itokawid 
candidates. Unfortunately, most of the physical information (e.g., bulk 
density, ablation coefficient, spectral data, etc.), are not given in the 
published data, except for the mass data. However, we could analyze 
corresponding mineral matters of the Itokawid candidates through their 
trajectory and atmospheric data instead, i.e., atmospheric density at 
meteor's luminous height, velocity, entry angle to Earth' atmosphere, and 
mass, using Ceplecha's (1988) classification. The Itokawid candidates are 
classified on the basis of evaluating the coefficients $K_B$ only for optical 
(Super-Schmidt and TV) meteors and $P_E$ and $A_L$ only for fireballs. In 
these evaluations, we adopted the atmospheric data from CIRA1965.

The results of classification are summarized in Table 3, where the column 
heads from left to right are as follows: object name, $V_\infty$: initial velocity in km 
s$^{-1}$, $\cos Z_R$: cosine of zenith distance of the apparent radiant, $H_B$ 
and $H_E$: beginning and end height of meteor in km, $m_\infty$: initial 
(photometric) mass in gram, $K_B$, $P_E$ and $A_L$, classified group, and 
classified mineral matter. It should be also noted that the $K_B$ coefficient 
for the TV meteors is adjusted adding 0.15 and the logarithm of the total 
luminosity of MORP172 for $A_L$ is 3.31 (in 0 mag sec).

\begin{table}[tdp]
\begin{center}
\rotatebox{90}{
\begin{minipage}{\textheight}
%\centering
\caption{Results of classification}
\vspace{3mm}
%\tiny
\begin{tabular}{lllccccccccc}
Object & $\> V_\infty$ & $\cos Z_R $ & $H_B$ & $H_E$ & $m_\infty$ & 
\multicolumn{5}{c}{Ceplecha's classification} \\
&&&&&& $K_{B}$ & $P_{E}$ & $A_{L}$ & group & mineral matter \\[1pt]
\hline\hline
H7220 & 12.8 & 0.97 & 78.8 && 0.085 & 7.62 &&& A & carbonaceous chondrite \\
H10330 & 11.4 & 0.82 & 77.6 && 0.027 & 7.61 &&& A & carbonaceous chondrite \\
HA10000 & 12.54 & 0.932 & 79.1 & 76.7 && 7.59 &&& A & carbonaceous chondrite \\
MORP172 & 12.5 & 0.967 &66.5 & 31.2 & 2200 & & $-4.57$ & 5.13 & I or II&  ordinary-carbonaceous \\
00504004 & 11.26 & 0.341&  83.1&&  0.045 &  7.53 &&& A & carbonaceous chondrite
\end{tabular}
\end{minipage}
}
\end{center}
\label{tab3}
\end{table}%

These evaluations indicate all the Itokawid candidates evidently belong to 
asteroidal meteors, as shown in Table 3: all the Super-Schmidt and TV 
meteors are classified in the range of group A as carbonaceous chondrite 
($7.30 \mathbin{\lower.3ex\hbox{$\buildrel<\over 
{\smash{\scriptstyle=}\vphantom{_x}}$}} K_B< 8.00$), while the fireball 
MORP172 is in group I as ordinary chondrite by $P_E$ classification, $-4.60 < 
P_E$, but in group II as carbonaceous chondrite by $A_L$ classification, 
$4.13 \mathbin{\lower.3ex\hbox{$\buildrel<\over 
{\smash{\scriptstyle=}\vphantom{_x}}$}} A_L< 5.36$. Hence, the 
classified mineral matter of MORP172 shows more similarity to Itokawa's 
surface composition of a LL chondrite analogue suggested by Binzel et al. 
(2001) than the four other meteors. Interestingly, according to Halliday et 
al (1989), MORP172 was observed as a likely meteorite dropping fireball 
with an estimated terminal mass of $\sim 1$ kg; this meteorite might have fallen, 
and still remains, somewhere near Edam, Canada.

\section{Conclusion}

As a result of a survey of the Itokawid meteors from among our multi-station 
optical meteor orbit data of $\sim 15000$ orbits, and applying the orbital 
similarity criteria, $D$-criteria, we could find five Itokawid meteor 
candidates. Two out of them, the meteor H7220 and the fireball MORP172, are 
probably associated with Itokawa, judging from their $D_{\rm SH}$ values. It is also 
notable that the secular near-invariants of MORP172 in the $U$-$\cos \theta $ 
coordinates match well with those of Itokawa: this suggests both objects 
have a strong dynamical relationship to each other.

The $K_B$, $P_E$ and $A_L$ evaluations indicate all the Itokawid candidates 
evidently belong to asteroidal meteors. MORP172 is the only one classified 
at ordinary/carbonaceous chondrite, the others being in carbonaceous 
chondrites. Hence, the classified mineral matter of MORP172 shows more 
similarity to Itokawa's surface composition of a LL chondrite analogue than 
the other candidates.

We conclude, on the basis of the investigations above, that the fireball, 
MORP172, is the strongest Itokawid candidate. However, if you ask whether 
there are Itokawid meteors or not, we will answer, ``Possible", since only a 
small sample like this is available so far. Therefore, more optical 
detections and analyses for the Itokawid meteors are desirable in the future 
work.

\section*{Acknowledgement.} 
We thank Messrs. C. Shimoda and S. Okumura for 
providing the JN monitoring data and Dr. P. Koten for offering unpublished 
trajectory data of the meteor, 00504004.

\end{document}